\documentstyle[12pt]{article}

\def\be{\begin{eqnarray}}
\def\ee{\end{eqnarray}}
\def\bq{\begin{equation}}
\def\eq{\end{equation}}
\def\ben{\begin{enumerate}}\def\een{\end{enumerate}}

\def\roughly#1{\mathrel{\raise.3ex\hbox{$#1$\kern-.75em%
\lower1ex\hbox{$\sim$}}}}

\begin{document}
\begin{titlepage}
\hfill{\large RPI-97-N117}
 
\hfill{\large FTUV 97/49}

\hfill{\large IFIC 97/80}
\vspace{.2cm}
\begin{center}
\ \\
{\large \bf On the Delta-Nucleon and Rho - Pi Splittings:}

{\bf A QCD-inspired Look in Free Hadrons versus Nuclei}
\ \\
\ \\
\vspace{1.0cm}
{Nimai C. Mukhopadhyay$^{(a,b)}$ and Vicente Vento$^{(a)}$}
\vskip 0.5cm

{\it (a) Departament de Fisica Te\`orica and Instituto
de F\'{\i}sica Corpuscular}

{\it Universitat de Val\`encia - Consejo Superior
de Investigaciones Cient\'{\i}ficas}

{\it 46100 Burjassot (Val\`encia), Spain}

{\it (b) Department of Physics, Applied Physics and Astronomy }

{\it Rensselaer Polytechnic Institute}

{\it Troy, NY 12180-3590, USA}

\today

\end{center}
\vskip 0.5cm
\centerline{\bf ABSTRACT}
\vskip 0.5cm
{Relationships between mass intervals for free hadrons
and in nuclei are studied in two theoretical approaches inspired by
QCD: naive quark model and skyrmion model, taking one example each
from mesons and baryons, that of pi-rho splitting in mesons, and
nucleon-Delta splitting in baryons. Possible
deconfinement effects in nuclei are examined.}

\vskip 0.9cm
\leftline{Pacs: 12.20.Fv, 12.39.Dc, 12.39.Jh, 12.40.Yx, 13.40.Hq }
\leftline{Keywords:  baryons, mesons, quark model, skyrmion model, nuclear
medium  }
\vspace{0.9cm}

{\tt
\leftline {nimai@nimai.phys.rpi.edu}
\leftline{vicente.vento@uv.es}}

\end{titlepage}

\section{Introduction}
\indent\indent Given the success of the standard model (SM) of the strong
and electroweak  interactions, all ingredients of understanding hadron
properties are in place at least through the energy scales set by the masses
of the W and Z bosons. The detailed structure studies of hadrons form a new
frontier of nuclear physics.

One aspect of this frontier is the study of {\em known} nuclear phenomena in
terms of the degrees of freedom of $QCD$, viz., quarks and gluons. In
the non-perturbative domain of $QCD$, the most rigorous attempts to
compare the properties of hadrons use lattice methods. There are,
however, numerous models inspired by $QCD$ to do the same in various
approximations. Two examples of the latter, to be used in this paper,
are the quark model \cite{rgg75,cl79} and the skyrmion model \cite{sk,br94}.

An important inference from $QCD$ is that the complex nucleus forms a
new vacuum, wherein new strong interaction phenomena are expected to
occur \cite{sh96}. While traditional nuclear methods may work extremely
well at low energy,
low temperature and/or nuclear density, $QCD$  provides new 
insights into novel nuclear phenomena,
many of which cannot be described by the traditional nuclear many-body
techniques. For example, formation of quark-gluon plasma is anticipated
in $QCD$, and is beyond the domain of traditional nuclear physics.

The purpose of this paper is to examine the splitting of the baryons and
mesons in free hadrons and in complex nuclei in two $QCD$-inspired
models, quark model and skyrmion model, taking the mass intervals
\be
\Delta(1232)& -& N , \nonumber\\
&& \label{md}\\
\rho(770)& - &\pi , \nonumber
\ee
$N$ being the nucleon and the $\pi$ the pi meson. In (\ref{md}), we shall
ignore the isospin breaking. For free baryons, the above interval will be
called $\Delta M$, and in nuclei $\Delta M ^*$; similarly $\Delta m$ and
$\Delta m^*$ for mesons respectively.

Reasons for taking these particular intervals are partly theoretical and
partly experimental. In the baryon case, there have been extensive
studies of the production and decay of $\Delta (1232)$, both off
nucleons and in nuclei, in the pi-meson factories \cite{pifac} and also
by photons \cite{framz}.
On the face of it, the intervals $\Delta M$ and $\Delta M^*$
seem to be  
not very hard to infer, 
$\Delta M$ being directly known, and the properties 
of $\Delta (1232)$ and nucleon in
nuclei can be used as inputs to determine 
$\Delta M^*$. Attempts to study higher energy
baryon resonances in nuclei have resulted in the discovery that many of
these resonances essentially disappear in nuclei heavier than the deuteron.
Thus, quantitative information on them are difficult to obtain.

The mesonic interval $\Delta m^*$ is not as well-known experimentally in
nuclei. Interpretation of $\Delta m$ and $\Delta m^*$ is also difficult,
given the complex property of the pion as a Goldstone boson.
Nevertheless, a discussion of the $\rho-\pi$ interval is theoretically
very interesting. In the naive quark model, it is simply related to the
$\Delta-N$ interval via the color hyperfine interaction \cite{cl79}.
Thus, we shall consider it in this paper, speculating about its value in
nuclei.

We note, at the outset, that traditional nuclear many-body methods (e.g.
\cite{lzmz}) can be used, with a great deal of success, to discuss
theoretically these intervals in nuclei. Thus, for $\Delta M^*$,
particle-hole and Delta-hole methods (\cite{lzmz}-\cite{oset}) are very
successful in explaining this interval in terms of complex nuclear
dynamics, {\em without invoking} $QCD$. Indeed, we cannot hope to even
approach such successes in the $QCD$-based approaches at present.
However, our modest goal is to offer different insights and connections
in a subject well-known in the traditional nuclear physics domain; we
believe that these are not easily gotten in the latter. Thus, the
intrinsic connection between the baryonic and mesonic intervals, due to
quark-gluon structure, is impossible to obtain in the traditional
many-body approaches not based on $QCD$, and we want to profit from the
former here.

\section{The Quark Model Approach}
\subsection{Free hadrons}
\indent\indent In the nonrelativistic quark model, the vector (V)-
pseudoscalar(PS) meson splitting is determined in terms of the color
magnetic hyperfine  interaction \cite{cl79}. Let us recall the model
of ref. \cite{halmar} for the ground-state hadron masses. The relevant
meson masses for determining the mesonic interval are given by
\bq
m(\pi) \approx 2m_u -\frac{3a}{m_u^2}, \label{pi}\\
\eq
\bq
m(\rho) \approx 2m_u +\frac{a}{m_u^2}, \label{rho}
\eq
where $a$ is the hyperfine constant defined by
\bq
a=\frac{8\pi}{9}\alpha_S |\Psi(0)|^2,
\label{qq}
\eq
$\alpha_S$ is the strong fine-structure constant and we are using, for the
up and down 
constituent quark masses, the approximation $m_u=m_d$, 
the quark density at the origin,
$\Psi(0)$, being not very model-dependent. From
Eqs. (\ref{pi}) and (\ref{rho}) one obtains for the mesonic interval
\bq
\Delta m \approx \frac{4a}{m_u^2}.
\eq
Analogously in the baryon case \cite{cl79},
\bq
M(N) \approx 3m -\frac{a'}{m_u^2}, \label{N}
\eq
\bq
M(\Delta) \approx 3m_u +\frac{a'}{m_u^2},\label{delta}
\eq
where $a'$ is the hyperfine constant for defined by
\bq
a'=\frac{2\pi}{3}\alpha_S |\Psi(0)|^2,
\label{qqq}
\eq
leading to
\bq
\Delta M  \approx \frac{2a'}{m_u^2}.
\eq
Choosing $a\approx a'$ we obtain
\bq
\Delta M \approx \frac{2a}{m_u^2} \approx \frac{1}{2}\Delta m
\approx 300\; \mbox{MeV} ,\label{dm}
\eq
which is well-satisfied by the data. This well-known result
is a reflection of the r\^ole of the color dynamics in the masses
of hadrons\cite{cl79,halmar}.

\subsection{Hadrons in nuclei}
\indent\indent We assume that the constituent quark mass and 
the hyperfine constant
{\em both} change due to altered $QCD$ vacuum in the nuclear medium
\cite{sh96}. Let $M_u$ and $A$ denote respectively the in-medium values
of these parameters, which are related to the free values by
\be
M_u& = & m_u + \delta,\label{M} \\
A & = & a + \varepsilon.\label{A}
\ee
We have, for nuclei,
\bq
m^*(\pi) \approx 2M_u -\frac{3A}{M_u^2}, \label{pi*}
\eq
\bq
m^*(\rho) \approx 2M_u +\frac{A}{M_u^2},\label{rho*}
\eq
\bq
M^*(N) \approx 3M_u -\frac{A}{M_u^2}, \label{N*}
\eq
\bq
M^*(\Delta) \approx 3M_u +\frac{A}{M_u^2},\label{delta*}
\eq

\subsection{Analysis of quark model parameters for nucleon and Delta}
\indent\indent The experimental situation in complex nuclei can be
summarized by the following \cite{pifac,framz}:
\bq
M^*(\Delta) \approx M(\Delta) \approx 1232 \pm 35 \mbox{MeV}, \label{md*}
\eq
\bq
M^*(N) < M(N)
\eq
We shall take the effective mass of the nucleon in nuclei to be
\bq
M^*(N) \approx 0.7 M(N), \label{mn*}
\eq
even as a bigger reduction of effective mass is indicated in the
Dirac-Brueckner-Hartree-Fock theories \cite{rimu96}. Thus we shall take
\bq
 M^*(\Delta) - M^*(N) \approx 576 \;\mbox{MeV}.\label{dm*}
\eq
From Eqs. (\ref{N}) and (\ref{delta}),
\bq
m_u \approx \frac{M(\Delta) + M(N)}{6} \approx 362
\;\mbox{MeV}.\label{mu}
\eq
From Eqs. (\ref{M}), (\ref{A}), (\ref{N*}), (\ref{delta*}) and (\ref{dm*}),
we obtain
\bq
\delta \approx -47 \; \mbox{MeV},
\eq
therefore $|\delta| << m_u$. Using this result and Eq. (\ref{dm}) we get
\bq
\frac{2A}{M_u^2} \approx \frac{2\varepsilon}{m_u^2} + \Delta M
\left(1-\frac{2\delta}{m_u}\right) \approx M(\Delta) - 0.7 M(N).
\eq
Thus
\bq
\frac{\varepsilon}{m_u^2} = 97 \; \mbox{MeV}. \label{eps}
\eq
A comparison of Eq. (\ref{dm}) and (\ref{eps}) indicates that $a$ and
$\varepsilon$ are comparable in magnitude given Eqs. (\ref{md*}) and
(\ref{mn*}).

We can summarize our effective baryon  analysis in comparing
free and nuclear properties: the effective quark mass in the nuclear
medium becomes
\bq
M_u = m_u + \delta \approx 315 \; \mbox{MeV},
\eq
{\em smaller} than the free quark mass of $362 \;\mbox{MeV}$
(Eq. (\ref{mu})). This drop in the quark mass is in accord with the
studies\cite{rimu96} in the Nambu-Jona Lasinio \cite{njl} model, where a
sharp drop of the quark mass is expected as a function of the nuclear
Fermi momentum. The hyperfine interaction term for the free baryons
\bq
\frac{a}{m_u^2} \approx \frac{M(\Delta) -M(N)}{2} \approx 147
\;\mbox{MeV},\label{a}
\eq
is less than
\bq
\frac{A}{m_u^2}  \approx  287 \;\mbox{MeV},
\eq
indicating that $A > a$. This is suggestive of an increase of the
effective coupling constant, $\alpha_S^{eff}$, in the nuclear medium and 
could be an indication of an interesting $QCD$ effect in the
properties of baryons in the medium, that of hadronic deconfinement.
The phenomenon could be investigated from the properties of baryonic
resonances in the nuclear medium, thus opening a complementary line of research
to that at the high energy scale, i.e.,$EMC$ type studies \cite{emc}, where
similar phenomena have been observed. How our non-perturbative
deconfinement scenario is related to the {\it nucleon swelling} discussion is
still under scrutiny.  

There are obviously many important implications of the phenomenon in
nuclei we have discussed above resulting in $M_u < m_u$ and $A>a$. These
involve various spectroscopic properties of baryons (and mesons) in
nuclei, some of which we shall discuss in a later section.

\subsection{Analysis of quark model parameters for pi and rho}
\indent\indent Requirement of self-consistency of the quark model demands
that the parameters obtained from the baryons and the mesons should broadly
agree with each other. This is generally the case, though the agreement
is not always perfect. An analysis of {\em all} mesons and baryons yield
such an agreement overall \cite{ik}. As we know, the cases for the Goldstone
bosons are always special and some disagreements involving them are not
surprising \cite{iach}. We shall examine below this issue.

If we compute the quark mass from the vector mesons \cite{halmar},
the value for the mass is  $390 \;\mbox{MeV}$, quite close
to the determination from baryons. However,
from the formulas for the free pions and rhos, Eq. (\ref{pi})
and (\ref{rho}) one obtains \cite{halmar}
\bq
m_u \approx \frac{m(\pi) + 3 m(\rho)}{8} \approx 306 \;\mbox{MeV}
\eq
This is considerably less than the value we have inferred earlier for
the baryons (Eq. (\ref{mu})). One could even use the above
equations to determine the quark mass from the
free rho and Delta masses (Eqs. (\ref{rho}) and (\ref{delta})) and
would obtain
\bq
m_u \approx M(\Delta) - m(\rho) \approx 462 \; \mbox{MeV}, \label{dr}
\eq
considerably more than the value from the baryons! These differences are
just the outcome of our oversimplified model, where no explicit
mechanism for confinement has been used. In this respect the model
assumes that confinement is only operative at large separations and
independent from quark masses and spins. However, it is clear that the
constituent masses themselves are a reflection of the confinement
mechanism \cite{cl79} and therefore the values of these masses are giving us
a consistency check. One should not mix baryons and mesons in the
process of determining the quark masses in this naive model.
Furthermore, given the pion having its Goldstone property, the
inequality $(m_u)_{meson} < (m_u)_{baryon}$ is not surprising,
as the contribution of quark masses is being reduced to satisfy the
Goldstone (low mass) property of the pion.
This discussion hints, however, at a likely interpretation for the $\delta$
parameter and its value. If we consider the quark mass as an energy
(momentum) parameter related to confinement \cite{cl79}, its value is
telling us that the confinement properties of the in-medium nucleon are
changing, and moreover {\it the negative sign implies} that a  deconfinement
process is taking place.  This explanation  agrees with the one hinted by
the $\varepsilon$ parameter.

The color hyperfine constants, obtained from the mesons and baryons under
consideration, are more stable than the quark masses themselves. Thus,
using Eqs. (\ref{pi}) and (\ref{rho}),
\bq
\left(\frac{a}{m_u^2}\right)_{meson} \approx \frac{m(\rho) - m(\pi)}
{4} \approx 158 \;\mbox{MeV},
\eq
in agreement with the value obtained in Eq. (\ref{a}) within a 7\%.

Using the baryon parameters for the free and nuclear cases, we can
conclude that
\bq
m^*(\rho) -m^*(\pi) > m(\rho) - m(\pi).
\eq
While $m^*(\rho)$ increases compared to $m(\rho)$, $m^*(\pi)$ drops
strongly in nuclei according to these considerations.
There is some evidence \cite{liu} of this behavior of  $m^*(\rho)$ on
the lattice, while the dropping of $m^*(\pi)$ is reminiscent of the
nuclear behavior as a {\em chiral filter} \cite{br} in the Brown-Rho
approach.

\section{The Skyrmion Approach}
\subsection{Free Baryons: the large $N$ scaling}
\indent\indent Let us examine the baryon interval problem from the
Skyrmion approach, where the crucial idea is the generalization of QCD
to $N$ colors, as was originaly proposed by 't Hooft \cite{ho} and Witten
\cite{wi}. In  large-$N$ $QCD$, there is a systematic expansion of
contributions to baryon properties in powers of $\frac{1}{N}$. The hope
here is, in Witten's words, {\it the $N = 3$ theory may be qualitatively
and quantitatively close to the large $N$ limit}. It is also possible to
connect this large $N$ limit to the results of the quark model in the
limit of large number of colors \cite{kp}.

Let us start by quoting some well-known results to set the scales of various
quantities of interest to us, obtained by Adkins, Nappi and Witten (ANW)
\cite{anw}. We begin with the Skyrme model Lagrangian \cite{sk,br94},
\bq
{\cal L} = \frac{F_\pi^2}{16} Tr(\partial_\mu U \partial^\mu U^+) +
\frac{1}{32e^2}Tr[(\partial_\mu U ) U^+,\partial_\nu U )
U^+]^2,\label{sk}
\eq
where $U$ is the usual SU(2) matrix of fields, $F_\pi$ is the pion decay
constant and the dimensionless parameter $e$ controls the strength of
the last term, the Skyrme term, needed to stabilize the soliton.

The scalings of $F_\pi$ and $e$ in number of colors, $N$, are given by
\bq
F_\pi^2 \sim N , \;\;\; \frac{1}{e^2} \sim N,
\eq
the symbol $\sim$ implying hereafter {\it scales as}.
(In the ANW analysis the values of the parameters used were
$e=5.45, \; F_\pi = 129\; \mbox{MeV}$.)

The skyrmion mass is given by
\bq
M = \frac{\pi}{2}\; F_\pi^2\; r_0\; {\cal I}({\cal A}),
\eq
where
\bq
{\cal I} = \int^\infty_{-\infty} d\tau \left[e^\tau \left(
\left(\frac{d\Theta}{dt}\right)^2 +  2(\sin \Theta)^2 \right) +
{\cal A} e^{-\tau} (\sin \Theta)^2 \left(
\left(\frac{d\Theta}{dt}\right)^2 +  \frac{1}{2}(\sin \Theta)^2 \right)
\right]
\eq
and
\bq
{\cal A} =\frac{4}{F_\pi^2 e^2 r_0^2},\label{aa}
\eq
$\tau=\ln {\frac{r}{r_0}}$, $\Theta$ is the so called chiral angle, and
$r_0$ is a 
suitable scale parameter 
\cite{jr}.
We get from ANW:
\bq
M=36.5 \;\frac{F_\pi}{e} \sim N.
\eq
The rotational energy from the quantization of the skyrmion is given by
the moment of inertia
\bq
\lambda = \frac{2}{3}\; \pi\;F_\pi^2 r_0^3 \; {\cal J}({\cal A}) \sim N.
\eq
Here ${\cal J}$ is the integral \cite{jr}
\bq
{\cal J} =  \int_{-\infty}^\infty d\tau
(\sin{\Theta})^2\left[1+{\cal A}\left[ \left(\frac{d\Theta}{dt}\right)^2
+\frac{(\sin{\Theta})^2}{r^2}\right]e^{-2\tau}\right]e^{-3\tau},
\label{int}
\eq
These details are useful for our later discussions. 

In terms of $\lambda$, the masses of the nucleon and the Delta as free
hadrons are given by
\bq
M(N) \approx M + \frac{3}{4}\;\frac{1}{2\lambda} \sim N + {\cal
O}(\frac{1}{N}),
\label{ns}
\eq
\bq
M(\Delta) \approx M + \frac{15}{4}\;\frac{1}{2\lambda} \sim N + {\cal
O}(\frac{1}{N}).
\label{ds}
\eq
Thus, the difference between them is given by
\bq
\Delta M = M(\Delta) -M(N) \sim \frac{1}{\lambda} \sim \frac{1}{N}.
\label{dms}
\eq
One could even optimistically anticipate that the Casimir and other
subleading corrections 
would drop out in the difference \cite{hs}.

\subsection{Baryons in the medium: scaling parameters}
\indent\indent We shall start from the known structure of the effective
Lagrangian at low energy and zero density (temperature), given by the theory
of skyrmions (\ref{sk}). Our discussion for complex nuclei will proceed as
follows: if we increase  the density (temperature) of the nucleus
(i.e., of the hadronic matter),  the properties of the vacuum  should change.
In the Skyrme model  modified for the nuclear medium the two relevant
parameters,
$F_\pi\; \mbox{and}\; e $ will change \cite{br}. We work in the chiral limit
($m_{\pi}= 0$),  which 
can still be taken in a situation affected  
by the increase in the hadronic
density
\cite{gb}  and/or temperature \cite{ga}. We focus 
here only on the nucleon and the
Delta in the nuclear medium.

In the Skyrme model, the baryonic interval is inversely proportional to
the moment of inertia, Eq.(\ref{dms}), which is given in terms of the
Skyrmion profile through a complicated integral, Eq. (\ref{int}), where
the dependence on the relevant parameters is complex.
We next prove that  the medium dependence of the moment of inertia
can be related to that of $g_A$. Thus, the model establishes a relation
between the baryonic interval and $g_A$. This relation has important
experimental implications which we shall analyze.

The density (temperature) scaling properties affect the skyrmion
profile, since, for the soliton configuration in the hedgehog form, one has
a unique solution for a given value of ${\cal A }$ defined in Eq. (\ref{aa}).
This solution satisfies the following scaling law \cite{jr,rho}
\bq
\Theta_{{\cal A'}}(\tau) =\Theta_{\cal A}\left(\tau -\frac{1}{2}\;
\ln{\frac{{\cal A'}}{{\cal A}}}\right).
\eq
For a conserved axial current, $g_A$ is determined by the asymptotic
behavior of the soliton solution,
\bq
\lim_{\tau \rightarrow \infty} \Theta(\tau) \rightarrow
\alpha({\cal A})e^{-2\tau},
\eq
and is given by the residue at the pion pole
\bq
g_A \sim 2\pi \alpha ({\cal A}) F_\pi^2r_0^2.
\eq
In principle, this relation holds only for the soliton. The physical
$g_A$ (see \cite{anw}) differs from this one by numerical color-dependent
factors, which will cancel in the ratios.

Using the previous set of equations, we obtain
\bq
M({\cal A} )\sim \sqrt{g_A}\; F_\pi\; \frac{{\cal I}({\cal A})} {\sqrt{\alpha
({\cal A})}},
\eq
and
\bq
\lambda ({\cal A}) \sim \frac{(\sqrt{g_A})^3}{F_\pi}\;
\frac{{\cal J} ({\cal A})}{(\sqrt{\alpha({\cal A})})^3},
\eq
The scaling is determined by
\bq
\frac{{\cal I} ({\cal A}')}{{\cal I} ({\cal A})} =
\left(\frac{{\cal J} ({\cal A}')}{{\cal J} ({\cal A})}\right)^{\frac{1}{3}} =
\left(\frac{\alpha ({\cal A}')}{\alpha ({\cal A})})\right)^{\frac{1}{2}},
\eq
and therefore all of the nuclear medium dependence will be given by
\bq
M({\cal A}) \sim \sqrt{g_A}\; F_\pi,
\eq
and
\bq
\lambda({\cal A}) \sim \frac{(\sqrt{g_A})^3}{F_\pi}.\label{ms}
\eq

The mass equation and the $N$ dependence suggest that the two
chosen parameters, $g_A$ and $F_\pi$,  scale according to
\bq
F_\pi \sim \sqrt{g_A}.
\eq
We now use this rule to come to the following scaling laws:
\bq
M \sim g_A,
\eq
and
\bq
\Delta M \sim M(\Delta) - M(N)  \sim \frac{1}{\lambda} \sim \frac{1}{g_A},
\eq
which lead, in the case of hadronic matter, to
\bq
M^* \sim g_A^*,
\eq
and
\bq
\Delta M^* \sim \frac{1}{g_A^*}.
\eq
This yields the scaling rule for the mass of the the nucleon and
Delta in the nuclear medium. In particular, Eqs. (\ref{md*}) and
(\ref{mn*})
imply
\bq
g_A > g_A^*.
\eq
Thus, the experimental results on the baryonic interval {\it imply, in the
leading order in $N$}, that
$g_A$ is quenched in the nuclear medium \cite{ncm}, a subject of
great topical interest in $QCD$.

We should point out that the experimental situation on the quenching 
of $g_A$ in the nuclear medium is far from settled, despite the long
history of the subject \cite{ncm}. Here we are pointing out a new angle to
this problem, via its novel connection to the baryonic interval.

\subsection{A numerical analysis of the Skyrmion scaling in the
Delta-Hole model}
\indent\indent According to the $\Delta$-hole model
(\cite{lzmz}-\cite{oset})
\bq
\delta g = \frac{g_A}{g_A^*} \approx 1 + \frac{4}{9}\;\left(\frac{f^*}{m_\pi}
\right)^2\;\frac{2}{M(\Delta)- M(N)}\: g_0 ' \;\rho,
\eq
where the ratio of the Watson-Lepore constants for the nucleon is given
by $\frac{f^*}{f} \approx 4$, $g_0 '  \approx 0.7 \pm 0.1$ is the
Landau-Migdal
 parameter and $\rho$ is the nuclear density. With these values,
\bq
\delta g \approx 1+ (0.273 \pm 0.039) \;\frac{\rho}{\rho_0},
\eq
$\rho_0$ is the nuclear matter density. We use this equation to show in
Tables 1  and 2 how the skyrmionic mass and the baryonic interval change with
density in this model.

\begin{table}[thb]
\caption[] {\small Skyrmion mass ratio $\frac{M^*}{M}$ as a function of
nuclear density
(normalized to nuclear matter density
) for three values of the Landau-Migdal parameter $g'_0 = 0.7 \pm 0.1$.}
\vskip .3cm
\begin{center}
\begin{tabular}{|c|c|c|c|c|c|c|}\hline
$\frac{\rho}{\rho_0}$&0.0&0.2&0.4&0.6&0.8&1.0\\ \hline
$0.6$&1.0&.955&.914&.877&.889&.810\\ \hline
$0.7$&1.0&.949&.902&.859&.821&.786\\ \hline
$0.8$&1.0&.942&.889&.842&.800&.762\\ \hline
\end{tabular}
\end{center}
\end{table}

\begin{table}[thb]
\caption[] {\small Baryonic interval ratio $\frac{\Delta M^*}{\Delta M}$
as a function of
nuclear density
(normalized to nuclear matter density) for three values of the Landau-Migdal
parameter $g'_0 = 0.7 \pm 0.1$.}
\vskip .3cm
\begin{center}
\begin{tabular}{|c|c|c|c|c|c|c|}\hline
$\frac{\rho}{\rho_0}$&0.0&0.2&0.4&0.6&0.8&1.0\\ \hline
$0.6$&1.0&1.047&1.094&1.140&1.187&1.234\\ \hline
$0.7$&1.0&1.054&1.109&1.164&1.218&1.273\\ \hline
$0.8$&1.0&1.062&1.125&1.187&1.250&1.312\\ \hline
\end{tabular}
\end{center}
\end{table}

It is interesting to note at this point that the results showed in 
Table 1 
allow us to explain in a qualitative fashion the observed behavior of the
masses in the medium. Once Eqs.(\ref{ns}) and (\ref{ds}) are scaled, they
lead
to 
expressions of the form
\bq
\sim g_A^* + \frac{\kappa}{g_A^*},
\eq
where $\kappa(\Delta) >> \kappa (\mbox{N})$. Thus, in the nucleon case,
 the first
term (which decreases with density) starts dominating, while in the $\Delta$
case, the diverse tendencies tend to cancel. Ultimately the second term
should
dominate and both masses should increase with density. In order to do a more
quantitative discussion one should include the missing terms, Casimir energy
and other subleading corrections \cite{hs}.
Unfortunately, we do not yet have reliable estimates of them.

\section{Conclusions}
\indent\indent We have analyzed the medium dependence of the properties
of hadrons in two different schemes with the same physical input: {\it
the nuclear medium dependence implies a change in the properties of the
vacuum
which translates into a scaling (density and/or temperature dependence)
of the free model parameters}.

In the quark model, the hadron mass intervals studied
are connected by the the color hyperfine interaction. 
Our analysis indicates that the
fundamental parameters of the model, $m_u$ and $a$, 
or, equivalently, the confinement scale
and
$\alpha_S$, change, signalling a deconfinement process in the medium.
This scenario leads naturally to a change of the
properties of the hadrons, e.g., a {\it renormalization} of the strong,
electromagnetic and weak vertices in the nuclear medium \footnote
{For example, the magnetic moments in the naive quark model are inversely
proportional to the quark masses. Therefore, the electromagnetic couplings
should be reduced in the medium. A similar analysis can be done for all other
interactions looking at the parameter dependence of the naive model
formulas\cite{cl79, halmar}.}. This is of great topical interest in our
understanding of nuclear $QCD$ properties.

The same reasoning has been applied to the skyrmion model. The change of
the baryonic interval from the free value to that in nuclei produces a
change in the parameters of the model $F_\pi$ and $e$. This scaling is
complex, since it enters into the profile function of the hedgehog. We
have  avoided a complex parameter fitting procedure 
in nuclear medium,
and have taken the implications of the
model at the qualitative level. We have thus obtained a
relation between two physical intervals of our interest. In our
analysis, the nuclear quenching of $g_A$ is directly related
to the quenching of
the moment of inertia and therefore to the growth of the baryonic
interval in the nuclear medium. From the physical point of view, the
quenching
of $g_A$ in nuclei is again a manifestation of deconfinement,
since the nucleon in
the nuclear medium is closer to a chirally symmetric phase. Recalling
Adkins {\it et al.} \cite{anw},  
observed
hadron characteristics are
calculated by means of the profile function and their explicit dependence
on the parameters. Thus all of them  have a calculable Large-N scaling
behavior. Moreover, many of these quantities can be directly related to the
moment of inertia and therefore their medium behavior can be obtained
directly
from the above expressions \footnote{For example $\mu_{N\Delta} \sim
\mu_p - \mu_n \sim \lambda$, thus this
much-studied  electromagnetic transition rate\cite{dav}
will be quenched in the nuclear medium. The strong coupling of nucleons,
Deltas and
pions are related to $g_A$ through appropriate Goldberger-Treiman
relation \cite{gt}  and
therefore also subject to quenching.}. Thus the change of the baryon mass
interval in the medium leads to a {\it renormalization} of the strong,
electromagnetic and weak vertices in nuclei, through the scaling of the
parameters and the dictates of large $N$ behavior.

In summary,
the present investigation has shown that  two different
models of hadronic intervals suggest one and the
same phenomenon: partial deconfinement in the  nuclear medium. It is
interesting that they 
are found from {\it different, but complementary,
perspectives, both inspired by QCD}.
The quark model
emphasizes approximate color dynamics, while the Skyrme model draws the
attention to chiral dynamics\footnote{A beautiful example of the
complementarity
of the two approaches is related to $g_A$. 
In the quark model with
three
colors, $g_A$ is independent of the mass interval and therefore remains
unquenched. In the Skyrme model, chirality bridges them and therefore
leads to the quenching in nuclei in the leading order of $N$.}. Moreover,
we have shown that the in-medium
properties change in a very specific way determined by the parameters of
the theory and thus, the {\it renormalization} of the interactions lead
to experimentally testable phenomena which open new avenues of $QCD$
exploration in nuclei not anticipated by the conventional many-body
theory.

More theoretical investigations  are necessary to
provide accurate experimental scenarios where the in-medium properties
advanced
in this investigation are envisaged. We are aware of one work
along these lines completed recently \cite{wfs}
in the context of the $\omega$-meson. There have been also  recent studies
of these issues in the context of the QCD sum rules for nuclei \cite{fst}.
Future works should clarify relations between them and QCD-inspired model
studies such as ours. It is apt to end here with a
plea \cite {fst}: ``Ask not what nuclear physics can do for
QCD. Ask what QCD can do for nuclear physics!''

\subsection*{Acknowledgments}
\indent\indent This work was mostly done while one of the authors (NCM)
was a Visiting Professor in the Department of Theoretical Physics in the
University of Valencia under the auspices of ``IBERDROLA de Ciencia y
Tecnologia", while partially supported by the U.S. Department of Energy.
He is grateful to Prof. E. Oset, for his generous support
as well as for the wonderful hospitality of the members of the Theory
Department. He is also thankful to Profs. E. Oset, S.K. Singh,
M. Strikman and M. Soyeur for various discussions on related issues. The
other author (VV) has been  supported in part by DGICYT-PB94-0080,
DGICYT-PB95-0134 and TMR programme of the European Commision
ERB FMRX-CT96-008. He is thankful to Profs. G.E. Brown and
M. Rho for clarifications on their work, and to Profs. P. Gonz\'alez, M.
Traini,
and Drs. F. Cano, A. Ferrando and S. Scopetta for discussions on 
related issues and a careful reading of the manuscript.
\pagebreak

\end{document}